\documentclass[runningheads,envcountsect,envcountsame]{llncs}
\usepackage{floatflt}

\usepackage{pslatex}
\usepackage[T1]{fontenc}
\usepackage{color}

\usepackage{floatflt}
\usepackage{amsmath}
\usepackage{amsfonts}
\usepackage{amssymb}
\usepackage{pslatex}
\usepackage{stmaryrd}

\newtheorem{fact}[theorem]{Fact}{\bfseries}{\itshape}

\def\ldouble{\langle\!\!\langle}
\def\rdouble{\rangle\!\!\rangle}

\newcommand{\lfp}{\mathrm{lfp}}
\newcommand{\gfp}{\mathrm{gfp}}
\newcommand{\Cu}{C_{\uparrow}}
\newcommand{\Cd}{C_{\downarrow}}
\newcommand{\Ku}{K_{\uparrow}}
\newcommand{\Kd}{K_{\downarrow}}
\newcommand\Pre{{\mathit{Pre}}}
\newcommand\Post{{\mathit{Post}}}
\newcommand\wPre{{\widetilde{\Pre}}}

\newcommand{\Nat}{{\mathbb{N}}} 
\renewcommand{\epsilon}{\varepsilon} 
\renewcommand{\setminus}{\smallsetminus} 
\renewcommand{\phi}{\varphi} 
\newcommand{\sat}{\models}      

\newcommand{\Cond}{{{\mathit{Cond}}}}
\newcommand{\Goal}{{{\mathit{Goal}}}}
\newcommand{\Final}{{{\mathit{Final}}}}
\newcommand{\Init}{\mathit{Init}}

\renewcommand{\c}{{c}} 
\newcommand{\C}{{\mathsf{C}}} 
\renewcommand{\L}{{\mathcal{L}}} 
\newcommand{\bfP}{{\textbf{P}}} 
\newcommand{\R}{{\mathcal{R}}}   

\newcommand{\Until}{{\mathsf{U}}}
\newcommand{\Next}{{\mathsf{X}}}
\newcommand{\Release}{{\mathsf{R}}}
\newcommand{\Lang}{{\mathsf{TL}}}

\newcommand{\Conf}{{\mathit{Conf}}}
\newcommand{\env}{{\textit{env}}}

\newcommand{\LTS}{{\mathit{LTS}}} 
\newcommand{\Mess}{{\mathsf{M}}} 
\newcommand{\op}{{\mathit{op}}}

\newcommand{\overto}[1]{\xrightarrow{\!\!#1\!\!}}
\newcommand{\step}[1]{\overto{#1}} 

\newcommand{\perf}{\mathrm{perf}}

\newcommand{\egdef}{\stackrel{\mbox{\begin{scriptsize}def\end{scriptsize}}}{=}}
\newcommand{\equivdef}{\stackrel{\mbox{\begin{scriptsize}def\end{scriptsize}}}{\Leftrightarrow}}

\newcommand{\sem}[1]{\llbracket #1 \rrbracket}
\newcommand{\rep}[1]{\sem{#1}}
\newcommand{\ex}[1]{\ldouble #1 \rdouble}

\begin{document}


\title{On computing fixpoints in well-structured regular model checking,
  with applications to lossy channel systems}
\titlerunning{On computing fixpoints, with applications to lossy channel systems}

\author{  C. Baier\inst{1}
\and
	  N. Bertrand\inst{2}
\and
	  Ph. Schnoebelen\inst{2}
}
\institute{
	  Universit{\"a}t Bonn, Institut f{\"u}r Informatik I, Germany
\and
	  LSV, ENS de Cachan \& CNRS, France
}

\maketitle




\begin{abstract}
We prove a general finite convergence theorem for ``upward-guarded''
fixpoint expressions over a well-quasi-ordered set. This has immediate
applications in regular model checking of well-structured systems, where a
main issue is the eventual convergence of fixpoint computations. In
particular, we are able to directly obtain several new decidability results
on lossy channel systems.
\end{abstract}



\section{Introduction}

\emph{Regular model
  checking}~\cite{kesten2001,bouajjani2000b,voronkov-csl03} is a
popular paradigm for the symbolic verification of models with infinite
state space.  It has been applied to varied families of systems
ranging from distributed algorithms and channel systems to hybrid
systems and programs handling dynamic data structures.

In regular model checking, one works with regular sets of
states and handles them via
finite descriptions, e.g., finite-state automata or regular expressions. Models
amenable to regular model checking are such that, when $S\subseteq \Conf$
is regular, then $\Post(S)$ (or $\Pre(S)$), the set of 1-step successors
(resp., predecessors), is again a regular set that
can be computed effectively from $S$. Since regular sets are
closed under Boolean operations, one can\footnote{
   Actually, such symbolic computations are possible with any class of
   representation closed under, and providing algorithms for, $\Pre$ or
   $\Post$, Boolean operations, vacuity~\cite{kesten2001,henzinger2005}.
} try to compute the reachability set
$\Post^*(\Init)$, as the
limit of the sequence
\begin{gather}
\tag{*}
\label{eq-post*}
S_0 := \Init;
\quad
S_1 := S_0\cup\Post(S_0);
\quad
\ldots
\quad
S_{n+1} := S_n\cup\Post(S_n);
\quad
\ldots
\end{gather}
Since equality of regular sets is decidable, the computation of
\eqref{eq-post*} can contain a test that detects if the limit is reached in
finite time, i.e., if $S_{n+1}=S_n$ for some $n\in\Nat$,

With infinite-state models, the main difficulty is \emph{convergence}. It
is very rare that a fixpoint computation like \eqref{eq-post*} converges in
finite time, and innovative techniques that try to compute directly, or guess and
check, or approximate the limit set $\Post^*(\Init)$, are currently under
active
scrutiny~\cite{BLW03,boigelot2003,bouajjani2004,habermehl04,BFLS-atva2005}.

\emph{Well-structured transition systems} (WSTS) are a generic family of
models for which the co-reachability set $\Pre^*(\Final)$ can be computed
symbolically with a backward-chaining version of
\eqref{eq-post*}~\cite{abdulla2000c,finkel98b}. For WSTS's, convergence of
the fixpoint computation is ensured by WQO theory: one handles
upward-closed sets, and increasing sequences of upward-closed sets always
converge in finite time when the underlying ordering is a
well-quasi-ordering (a WQO), as is the case with WSTS's.

Computing $\Pre^*(\Final)$ for reachability analysis is just a special
case of fixpoint computation. When dealing with richer temporal
properties, one is interested in more complex fixpoints. E.g., the set
of states satisfying the CTL formula $\exists[\Cond\Until\Goal]$ is
definable via a least-fixpoint expression: $\mu
X.\Goal\cup(\Cond\cap\Pre(X))$. For game-theoretic properties, similar
fixpoints are involved. E.g., the states from which the first player
in a turn-based game can enforce reaching a goal is given by $\mu
X.\Goal\cup\Pre(\overline{\Pre(\overline{X})})$.

\paragraph*{\bf Our contribution.}
In this paper, we define a notion of $\mu$-expressions where recursion is
guarded by upward-closure operators, and give a general finite convergence
theorem for all such expressions. The consequence is that these fixpoint
expressions can be evaluated symbolically by an iterative procedure. The
guarded fragment we isolate is very relevant for the verification of
well-structured transition systems as we demonstrate by providing several
new decidability results on channel systems.

\paragraph*{\bf Related work.}
Henzinger \textit{et al.} give general conditions for the convergence of
fixpoints computations for temporal~\cite{henzinger2005} or
game-theoretic~\cite{alfaro2001b} properties, but the underlying framework
(finite quotients) is different and has different applications (timed and
hybrid systems). Our applications to well-structured transition systems
generalize results from ~\cite{abdulla2003b,raskin03,raskin05,KucSch-TCS}
that rely on more ad-hoc finite convergence lemmas.




\section{A guarded mu-calculus}
\label{sec-guarded-mucalculus}

We assume basic understanding of $\mu$-calculi techniques (otherwise
see~\cite{arnold2001}) and of well-quasi-ordering (WQO) theory (otherwise
see~\cite{milner85,kruskal72}, or simply~\cite[sect.~2.1]{finkel98b}).

Let $(W,\sqsubseteq)$ be a well-quasi-ordered set. A subset $V$ of $W$ is
\emph{upward-closed} if $w\in V$ whenever $v\sqsubseteq w$ for some $v\in
V$. From WQO theory, we mostly need the following result:
\begin{fact}[Finite convergence]
\label{fact-fc}
If $V_0\subseteq V_1\subseteq V_2\subseteq \cdots$ is an infinite
increasing sequence of upward-closed subsets of $W$, then
for some index $k\in\Nat$, $\bigcup_{i\in\Nat}V_i =V_k$.
\end{fact}

The \emph{upward-closure} of $V\subseteq W$, denoted $\Cu(V)$, is the
smallest upward-closed set that contains $V$. The \emph{upward-kernel} of
$V$, denoted $\Ku(V)$, is the largest upward-closed set included in $V$.
There are symmetric notions of \emph{downward-closed} subset of $W$, of
\emph{downward-closure}, $\Cd(V)$, and of \emph{downward-kernel}, $\Kd(V)$,
of $V$. The complement of an upward-closed subset is downward-closed.
Observe that $\Cu(V)=V=\Ku(V)$ iff $V$ is upward-closed, and that $\Cu$ and
$\Kd$ (resp., $\Cd$ and $\Ku$) are dual:
\begin{xalignat}{2}
\label{eq-Cu-duality}
        W\setminus \Ku(V) &= \Cd(W\setminus V),
        &
        W\setminus \Kd(V) &= \Cu(W\setminus V).
\end{xalignat}

\paragraph*{\bf Monotonic region algebra.}
In symbolic model-checking, a \emph{region algebra} is a family of
sets of states (subsets of $W$) that is closed under Boolean and other
operators like images or inverse images~\cite{henzinger2005}.

Here we consider regions generated by a family $O=\{o_1,o_2,\ldots\}$ of
(monotonic) operators. By a $k$-ary \emph{operator}, we mean a monotonic
mapping $o:(2^W)^k\to 2^W$ that associates a subset
$o(V_1,\ldots,V_k)\subseteq W$ with any $k$ subsets $V_1,\ldots,V_k$.
Monotonicity means that $o(V_1,\ldots,V_k)\subseteq o(V'_1,\ldots,V'_k)$
when $V_i\subseteq V'_i$ for $i=1,\ldots,k$. We allow nullary operators,
i.e., fixed subsets of $W$. Finally, we require that $O$ contains at least
four special unary operators: $\Cu$, $\Cd$, $\Ku$, $\Kd$, and two special
nullary operators: $\emptyset$ and $W$.

The \emph{region algebra generated by $O$}, denoted with $\R_O$, or simply
$\R$, is the set of all the subsets of $W$, called \emph{regions}, that can
be obtained by applying operators from $O$ on already constructed regions,
starting with nullary operators. Equivalently, $\R$ is the least subset of
$2^W$ that is closed under $O$.

We say the region algebra generated by $O$ is \emph{effective} if there are
algorithms implementing the operators in $O$ and an effective membership
algorithm saying whether $w\in R$ for some $w\in W$ and some region $R\in\R_O$.
Such effectiveness assumptions presuppose a finitary encoding of regions
and elements of $W$: if there are several possible encodings for a same
region, we assume an effective equality test.

\paragraph*{\bf Extending the region algebra with fixpoints.}
Let $\chi = \{X_1,X_2,\cdots\}$ be a countable set of variables.
$L_\mu(W,\sqsubseteq,O)$, or shortly $L_\mu$ when $(W,\sqsubseteq)$ and $O$
are clear from the context, is the set of \emph{$O$-terms with least and
greatest fixpoints} given by the following abstract syntax:
\begin{gather*}
L_\mu \ni \phi,\psi  \ ::=
               o(\phi_1,\ldots,\phi_k)          \ \big| \
               X                                \ \big| \
               \mu X.\phi                       \ \big| \
               \nu X. \phi                      \ \big| \
               \Cu(\phi)                        \ \big| \
               \Cd(\phi)                        \ \big| \
               \Ku(\phi)                        \ \big| \
               \Kd(\phi)
\end{gather*}
where $X$ runs over variables from $\chi$, and $o$ over operators from $O$.
$\mu X.\phi$ and $\nu X.\phi$ are fixpoint expressions. Free and bound
occurrences of variables are defined as usual. We assume that no variable
has both bound and free occurrences in some $\phi$, and that no two
fixpoint subterms bind the same variable: this can always be ensured by
renaming bound variables. (The abstract syntax for $L_\mu$ could be shorter
but we wanted to stress that $\Cu$, $\Cd$, $\Ku$, and $\Kd$ are required to
be present in $O$.)

The meaning of $L_\mu$ terms is as expected: an \emph{environment} is a
mapping $\env : \chi \to 2^W$ that interprets each variable $X \in \chi$ as
a subset of $W$. Given $\env$, a term $\phi\in L_\mu$ denotes a subset of
$W$, written $\rep{\phi}_\env$ and defined by induction on the structure of
$\phi$:
\begin{xalignat*}{2}
\rep{X}_\env & \egdef  \env(X)
&
\rep{o(\phi_1,\ldots,\phi_k)}_{\env} & \egdef
   o(\rep{\phi_1}_\env,\ldots,\rep{\phi_k}_\env)
\\
\rep{\Cu( \phi)}_\env & \egdef
   \Cu\bigl( \rep{\phi}_{\env} \bigr)
&
\rep{\Cd( \phi)}_\env & \egdef
   \Cd (\rep{\phi}_{\env})
\\
\rep{\Ku (\phi)}_\env & \egdef
   \Ku( \rep{\phi}_{\env})
&
\rep{\Kd (\phi)}_\env & \egdef
   \Kd\bigl( \rep{\phi}_{\env}\bigr)
\\
\rep{\mu X. \phi}_\env & \egdef  \lfp \bigl(\Omega[\phi,X,\env]
\bigr)
\!\!\!\!\!\!\!\!\!\!\!\!\!\!\!\!\!\!\!\!\!\!\!\!\!\!
\!\!\!\!\!\!\!\!\!\!\!\!\!\!\!\!\!\!\!\!\!\!\!\!\!\!
\!\!\!\!\!\!\!\!\!\!\!\!\!\!\!\!\!\!\!\!\!\!\!\!\!\!
&
\rep{\nu X. \phi}_\env & \egdef  \gfp \bigl(\Omega[\phi,X,\env]
\bigr)
\!\!\!\!\!\!\!\!\!\!\!\!\!\!\!\!\!\!\!\!\!\!\!\!\!\!
\!\!\!\!\!\!\!\!\!\!\!\!\!\!\!\!\!\!\!\!\!\!\!\!\!\!
\!\!\!\!\!\!\!\!\!\!\!\!\!\!\!\!\!\!\!\!\!\!\!\!\!\!
\end{xalignat*}
where $\Omega[\phi,X,\env] : 2^W \to 2^W$ is a unary operator defined by
$\Omega[\phi,X,\env](V) \egdef \rep{\phi}_{\env[X:= V]}$, using the
standard variant notation ``$\env[X:=V]$'' for the environment that agrees
with $\env$ everywhere except on $X$ where it returns $V$. As usual,
$\rep{\phi}_\env$ does not depend on $\env(X)$ if $X$ is not free in
$\phi$, so that we may shortly write $\rep{\phi}$ when $\phi$ is a closed
term, i.e., a term with no free variables.

We recall that the semantics of the fixpoint terms is well-defined since,
for every $\phi$, $X$ and $\env$, $\Omega[\phi,X,\env]$ is monotonic (and
since $(2^W,\subseteq)$ is a complete lattice). Moreover, if $\env$ and
$\env'$ are such that $\env(X)\subseteq\env'(X)$ for all $X\in\chi$,
shortly written $\env\subseteq\env'$, then $\lfp(\Omega[\phi,X,\env])
\subseteq \lfp(\Omega[\phi,X,\env'])$ and $\gfp(\Omega[\phi,X,\env])
\subseteq \gfp(\Omega[\phi,X,\env'])$.

\begin{definition}[Upward- and downward-guardedness]
\begin{enumerate}
\item A variable $X$ is \emph{upward-guarded} in $\phi$ if all free
  occurrences of $X$ in $\phi$ are in the scope of either a $\Cu$ or a
  $\Ku$ operator, i.e., appear in a subterm of the form $\Cu(\psi)$ or $\Ku
  (\psi)$.
\item Dually, $X$ is \emph{downward-guarded} in $\phi$ if all its
  free occurrences are in the scope of a
  $\Cd$ or a $\Kd$ operator.
\item A term $\phi$ is \emph{guarded} if all its least-fixpoint subterms
$\mu X.\psi$ have $X$ upward-guarded in $\psi$, and all its
  greatest-fixpoint subterms $\nu X.\psi$ have $X$ downward-guarded in $\psi$.
\end{enumerate}
\end{definition}

Given some $\phi$, $X$ and $\env$, the approximants of
$\lfp(\Omega[\phi,X,\env])$ are given by the sequence $(M_i)_{i\in\Nat}$ of
subsets of $W$ defined inductively with $M_0= \emptyset$ and $M_{i+1} =
\rep{\phi}_{\env[X:=M_i]}$. Monotonicity yields
\begin{gather}
\label{eq-mon}
M_0 \subseteq M_1 \subseteq M_2 \subseteq \cdots \subseteq \lfp(\Omega[\phi,X,\env]).
\end{gather}
Similarly we define $(N_i)_{i\in\Nat}$ by $N_0 = W$ and $N_{i+1} =
\rep{\phi}_{\env[X:=N_i]}$, so that
\begin{gather}
N_0 \supseteq N_1 \supseteq N_2 \supseteq \cdots \supseteq \gfp(\Omega[\phi,X,\env]).
\end{gather}

\begin{lemma}[Finite convergence of approximants]
\label{lem-finite-approx}
If $X$ is upward-guarded in $\phi$, then there exists an index $k\in\Nat$ such that
\begin{gather}
\label{eq-pfa}
\rep{\mu X.\phi}_\env =  M_k = M_{k+1} = M_{k+2} = \ldots
\end{gather}
Dually, if $X$ is downward-guarded in $\phi$, then there exists a $k'\in\Nat$
such that
\begin{gather}
\rep{\nu X.\phi}_\env = N_{k'} = N_{k'+1} = N_{k'+2} = \ldots
\end{gather}
\end{lemma}
\begin{proof}
We only prove the first half since the other half is dual. Let
$\psi_1,\ldots,\psi_m$ be the maximal subterms of $\phi$ that are
immediately under the scope of a $\Cu$ or a $\Ku$ operator. Then $\phi$ can
be decomposed under the form
\begin{gather*}
\phi \:\equiv\: \Phi(\Uparrow \psi_1,\ldots,\Uparrow \psi_m)
\end{gather*}
where the context $\Phi(Y_1,\ldots,Y_m)$ uses fresh variables
$Y_1,\ldots,Y_m$ to be substituted in, and where $\Uparrow\psi_i$ is either
$\Cu(\psi_i)$ or $\Ku(\psi_i)$, depending on how $\psi_i$ appears in
$\phi$. In either case, and for any environment $\env'$, the set
$\rep{\Uparrow \psi_i}_{\env'}$ is upward-closed.

For $V_1,\ldots,V_m \subseteq W$ we shortly write
$\rep{\Phi}(V_1,\ldots,V_m)$ for $\rep{\Phi}_{\env[Y_1 :=
V_1,\ldots,Y_m:=V_m]}$. Since $X$ is upward-guarded in $\phi$, it has no
occurrence in $\Phi$, only in the $\psi_i$'s, so that
\begin{align*}
        M_{i+1} = \rep{\phi}_{\env[X:=M_i]} & =
         \rep{\Phi}(\rep{\Uparrow \psi_1}_{\env[X:=M_i]},\ldots,
            \rep{\Uparrow \psi_m}_{\env[X:=M_i]}) \\
        & =
         \rep{\Phi}(L_{i,1},\ldots, L_{i,m})
\end{align*}
writing $L_{i,j}$ for $\rep{\Uparrow\psi_j}_{\env[X:=M_i]}$. From
$M_0\subseteq M_1\subseteq M_2 \subseteq \cdots$, we deduce $L_{0,j}
\subseteq L_{1,j} \subseteq L_{2,j} \subseteq \cdots$ Since $\Ku$ and $\Cu$
return upward-closed sets, the $L_{i,j}$'s are upward-closed subsets of
$W$. For all $j=1,\ldots,m$, Fact~\ref{fact-fc} implies that there is an
index $k_j$ such that $L_{i,j} = L_{k_j,j}$ for all $i \geq k_j$. Picking
$K=\max(k_1,\ldots,k_j)$ gives for any $i\geq K$
\[
M_{i+1} = \rep{\Phi}(L_{i,1},\ldots, L_{i,m})
        = \rep{\Phi}(L_{k_1,1},\ldots,L_{k_m,m})
        = \rep{\Phi}(L_{K,1},\ldots,L_{K,m})
        = M_{K+1}.
\]
Thus, $\bigcup_{i\in\Nat}M_i=M_{K+1}=M_{K+2}$ and $M_{K+1}$ is a fixpoint
of $\Omega[\phi,X,\env]$, hence the least one thanks to \eqref{eq-mon}.
Picking $k=K+1$ satisfies \eqref{eq-pfa}.
\qed
\end{proof}

\paragraph*{\bf Regions with guarded fixpoints.}
We can now prove our main result: subsets defined by $L_\mu$ terms are
regions (and can be computed effectively if the underlying region
algebra is effective).

By a \emph{region-environment} we mean an environment $\env : \chi \to \R$
that associates regions with variables. If $\env$ is a region-environment,
and $\phi$ has only free variables, i.e., has no fixpoints subterms, then
$\rep{\phi}_\env$ is a region.

\begin{theorem}
\label{th-guarded-effective}
If $\phi\in L_\mu$ is guarded and $\env$ is a region-environment then
$\rep{\phi}_\env$ is a region. Furthermore, if the region algebra is
effective, then $\rep{\phi}_\env$ can be computed effectively from $\phi$
and $\env$.
\end{theorem}
\begin{proof}
By structural induction on the structure of $\phi$. If $\phi=o()$ is a
nullary operator, the result holds by definition of the region algebra. If
$\phi = o(\phi_1,\cdots,\phi_k)$, the $\rep{\phi_i}_\env$'s are
(effectively) regions by induction hypothesis, so that $\rep{\phi}_\env$ is
an (effective) region too by definition. In particular, this argument
applies when $o$ is a nullary operator, or is one of the unary operators we
singled out: $\Cu$, $\Cd$, $\Ku$, and $\Kd$.

If $\phi = \mu X. \psi$, we can apply Lemma~\ref{lem-finite-approx} after
we have proved that each one of the approximants $M_0, M_1, M_2, \ldots$,
of $\rep{\phi}_\env$ are regions. In particular, $M_0=\emptyset$ is a
region, and if $M_i$ is a region, then $M_{i+1}=\rep{\psi}_{\env[X:=M_i]}$
is one too, since $\env'=\env[X:=M_i]$ is a region-environment, and since
by induction hypothesis $\rep{\psi}_{\env'}$ is a region when $\env'$ is a
region-environment. When $\R_O$ is effective, the $M_i$ can be computed
effectively, and one can detect when $M_k=M_{k+1}$ since region equality is
decidable by definition. Then $\rep{\phi}_\env = M_k$ can be computed
effectively. Finally, the case where $\phi=\nu X.\psi$ is dual.
\qed
\end{proof}
\begin{corollary}[Decidability for guarded $\L_\mu$ properties]
The following problems are decidable for effective
  monotonic region algebras:
\begin{description}
\item[Model-checking:]
``Does  $w \in \rep{\phi}$?'' for a $w\in
 W$ and a closed and guarded $\phi\in L_\mu$.
\item[Satisfiability:]
``Is  $\rep{\phi}$ non-empty?'' for a closed and guarded $\phi\in L_\mu$.
\item[Universality:]
``Does  $\rep{\phi}=W$?'' for a closed and guarded $\phi\in L_\mu$.
\end{description}
\end{corollary}

\paragraph*{\bf A region algebra of regular languages.}
Consider $W=\Sigma^*$, the set of finite words over some finite alphabet
$\Sigma$. The \emph{subword ordering}, defined by ``$u\sqsubseteq v$ iff $u$
can be obtained by erasing some letters from $v$'', is a WQO (Higman's
Lemma). Regular languages over $\Sigma$ are a natural choice for regions:
observe that the closure operators $\Cu$ and $\Cd$ preserve regularity and
have effective implementations.\footnote{
  From a FSA for $R$, one obtains a FSA for $\Cu(R)$ simply by adding loops
  $q\step{a}q$ on all states $q$ of the FSA and for all letters
  $a\in\Sigma$. A FSA for $\Cd(R)$ is obtained by adding
  $\epsilon$-transitions $q\step{\epsilon}q'$ whenever there is a
  $q\step{a}q'$. From this, $\Ku$ and $\Kd$ can be implemented using
  \eqref{eq-Cu-duality}.
} Natural operators to be considered in $O$ are $\cup$ (union) and $\cap$
(intersection). However, any operation on languages that is monotonic,
preserve regularity, and has an effective implementation on regular
languages can be added. This includes concatenation (denoted $R.R'$),
star-closure (denote $R^*$), left- and right-residuals
($R^{-1}R'\egdef\{v~|~\exists u\in R,uv\in R'\}$), shuffle product (denoted
$R \parallel R'$), reverse (denoted $\overleftarrow{R}$), conjugacy
($\widetilde{R}\egdef\{vu~|~uv\in R\}$), homomorphic and
inverse-homomorphic images, and many more~\cite{perrin90}. Complementation
is not allowed in $O$ (it is not monotonic) but the duals of all
above-mentioned operators can be included in $O$ (without compromising
effectiveness) so that, for all practical purposes, complement can be used
with the restriction that bound variables in $L_\mu$ terms are under an
even number of complementations.

An application of Theorem~\ref{th-guarded-effective} is that, if $R_1$ and
$R_2$ are regular languages, then the language defined as $\mu X.\nu
Y.\bigl(\Ku\bigl[R_1 \parallel (X^*\cap \Cd(Y^{-1}\overleftarrow{X}\cap
X^{-1}R_2))\bigr]\bigr)$ is regular and a finite representation for it
(e.g., a regular expression or a minimal DFA) can be constructed from $R_1$
and $R_2$.



\section{Verification of lossy channel systems}
\label{sec-verif-LCS}

Theorem~\ref{th-guarded-effective} has several applications for regular
model checking of lossy channel systems~\cite{abdulla96b} (LCS) and other
families of well-structured systems~\cite{abdulla2000c,finkel98b}. In the
rest of this paper we concentrate on LCS's.

\subsection{Channel systems, perfect and lossy}
A channel system is a tuple $\L= (Q,\C,\Mess,\Delta)$ consisting of a
finite set $Q=\{p,q,\ldots\}$ of \emph{locations}, a finite set
$\C=\{c,\ldots\}$ of \emph{channels}, a finite \emph{message alphabet}
$\Mess=\{m,\ldots\}$ and a finite set $\Delta=\{\delta,\ldots\}$ of
\emph{transition rules}. Each transition rule has the form $q \step{\op} p$
where $\op$ is an \emph{operation}: $\c!m$ (sending message $m \in \Mess$
along channel $\c \in \C$), $\c?m$ (receiving message $m$ from channel
$\c$), or $\surd$ (an internal action to some process, no I/O-operation).

\paragraph*{\bf Operational semantics.}
Let $\L= (Q,\C,\Mess,\Delta)$ be a channel system. A \emph{configuration}
(also, a \emph{state}) is a pair $\sigma = (q,w)$ where $q \in Q$ is a
location and $w : \C \to \Mess^*$ is a channel valuation that associates
with any channel its content (a sequence of messages). The set $Q\times
{\Mess^*}^\C$ of all configurations is denoted by
$\Conf=\{\sigma,\rho,\ldots\}$. For a subset $V$ of $\Conf$, we let
$\overline{V} \egdef \Conf \setminus V$.

Steps between configurations are as expected. Formally, $\sigma = (q,w)$
leads to $\sigma' = (q',w')$ by firing $\delta = p \step{\op} r$, denoted
$\sigma \step{\delta}_{\perf} \sigma'$, if and only if $q=p$, $q'=r$ and
$w'$ is obtained from $w$ by the effect of $\op$ (the ``$\perf$'' subscripts
emphasizes that the step is perfect: without losses). Precisely,
$w'(c)=w(c)$ for all channels $c$ that are not touched upon by $\op$, and
\[
	w'(c) = \begin{cases}
		  w(c) m & \textrm{ if } \op = c!m,\\
		  m^{-1}w(c) & \textrm{ if } \op = c?m.
		\end{cases}
\]
Thus, when $\op=c?m$, $w'$ is only defined if $w(c)$ starts with $m$ and
indeed this is the intended condition for firing $\delta$. Whenever $\sigma
\step{\delta} \rho$ for some $\rho$, we say that $\delta$ is \emph{enabled}
in $\sigma$, written $\delta \in \Delta(\sigma)$.

Below we restrict our attention to LCS's where from each $q\in Q$ there is
at least one rule $q\step{\op}p$ in $\Delta$ where $\op$ is not a receiving
action: this ensures that the LCS has no deadlock states and simplifies
many technical details without losing any generality.

\paragraph*{\bf Lossy systems.}
In \emph{lossy} channel systems, losing messages is formalized via the
subword ordering, extended from $\Mess^*$ to $\Conf$: $(q,w) \sqsubseteq
(q',w')$ if $q=q'$ and $w(c) \sqsubseteq w'(c)$ for all channels $c \in
\C$.

A (possibly lossy) step in the LCS is made of a perfect step followed by
arbitrary losses:\footnote{
\label{STEP-fn}
   Note that, with this definition, message losses only occur \emph{after}
   steps (thus, not in the initial configuration). The usual definition
   allows arbitrary losses before and after a step. There is no essential
   semantical difference between these two ways of grouping atomic events
   into single ``steps''. The usual definition is technically smoother when
   LCS's are viewed as nondeterministic systems, but becomes unnatural in
   situations where several adversarial processes compete, e.g., in
   probabilistic LCS's~\cite{BBS-acmtocl06} or other game-theoretical
   settings we explore in sections~\ref{sec-turn-based}
   and~\ref{sec-prob-lcs}.
} formally, we write $\sigma \step{\delta}\rho$ whenever there is a perfect
step $\sigma \step{\delta}_{\perf} \sigma'$ such that $\rho \sqsubseteq
\sigma'$. This gives rise to a labeled transition system $\LTS_\L \egdef
(\Conf,\Delta,\to)$, that can be given a WSTS structure by the following
relation: $\sigma\preceq \rho\;\equivdef\;
\sigma\sqsubseteq\rho\:\cap\:\Delta(\sigma)=\Delta(\rho)$. Both
$\sqsubseteq$ and $\preceq$ turns $\Conf$ into a WQO.
\begin{remark}
From now on we assume for the sake of simplicity that
$(\Conf,\sqsubseteq)$ is the WQO on which $L_\mu$ is defined. All results
could be strengthened using $(\Conf,\preceq)$.
\qed
\end{remark}
Following standard notations for transition systems $(\Conf,\Delta,\to)$
labeled over some $\Delta$, we write
$\Pre[\delta](\sigma)\egdef\{\rho\in\Conf~|~\rho\step{\delta}\sigma\}$ for
the set of predecessors via $\delta$ of $\sigma$ in $\L$. Then
$\Pre(\sigma) \egdef \bigcup_{\delta \in \Delta} \Pre[\delta](\sigma)$ has
all 1-step predecessors of $\sigma$, and $\Pre(V) = \bigcup_{\sigma \in V}
\Pre(\sigma)$ has all 1-step predecessors of states in $V$. The dual
$\wPre$ of $\Pre$ is defined by $\wPre(V)
=\overline{\Pre(\overline{V})}$. Thus $\sigma \in \wPre(V)$ iff
all 1-step successors of $\sigma$ are in $V$ (this includes the case where
$\sigma$ is a deadlock state).

Seen as unary operators on $2^\Conf$, both $\Pre$ and $\wPre$ are
monotonic and even continuous for all transition systems~\cite{sifakis82}.
For LCS's, the following lemma states that $\Pre$ is compatible with the
WQO on states, which will play a crucial role later when we want to show
that some $L_\mu$ term is guarded.
\begin{lemma}
\label{lem-pre-guarded}
Let $V\subseteq\Conf$ in the transition system $\LTS_\L$ associated with a
LCS $\L$. Then $\Pre(V) = \Pre(\Cu(V))$ and $\wPre(V) =
\wPre(\Kd(V))$.
\end{lemma}
\begin{proof}
$V\subseteq\Cu(V)$ implies $\Pre(V) \subseteq \Pre(\Cu(V))$. Now
$\sigma\in\Pre(\Cu(V))$ implies that $\sigma\step{}\rho\sqsupseteq\rho'$
for some $\rho'\in V$. But then $\sigma\step{}\rho'$ by definition of lossy
steps and $\sigma\in\Pre(V)$. The second equality is dual.
\qed
\end{proof}

\paragraph*{\bf An effective region algebra for LCS's.}
We are now ready to apply the framework of
section~\ref{sec-guarded-mucalculus} to regular model checking of lossy
channel systems. Assume $\L= (Q,\C,\Mess,\Delta)$ is a given LCS. A region
$R\in\R$ is any ``regular'' subset of $\Conf$. More formally, it is any set
$R\subseteq\Conf$ that can be written under the form
\[
	R = \sum_{i \in I} (q_i,R_{i}^1,\ldots,R_{i}^{|\C|})
\]
where $I$ is a \emph{finite} index set, the $q_i$'s are locations from $Q$,
and each $R_{i}^j$ is a regular language on alphabet $\Mess$. The notation
has obvious interpretation, with summation denoting set union (the empty
sum is denoted $\emptyset$). We are not more precise on how such regions
could be effectively represented (see~\cite{abdulla2001b}), but they could be handled as, e.g.,
regular expressions or FSAs over the extended alphabet $\Mess\cup
Q\cup\{\mathtt{'('},\mathtt{')'},\mathtt{','}\}$.

The set $O$ of operators includes union, intersection, $\Cu$, $\Cd$, $\Ku$,
$\Kd$: these are monotonic, regularity-preserving, and effective operators
as explained in our example at the end of
section~\ref{sec-guarded-mucalculus}. Operators specific to regular
model-checking are $\Pre$ and $\wPre$. That they are
regularity-preserving and effective is better seen by first looking at the
special case of perfect steps:
\begin{align*}
\Pre_{\perf}[p \step{c_i?m} r](q,R_p^1, \cdots, R_p^{|\C|}) &=
\begin{cases}
  (p,R_p^1, \ldots, R_p^{i-1}, m R_p^i, R_p^{i+1}, \ldots, R_p^{|\C|}) & \textrm{ if } q=r,\\
  \emptyset & \textrm{ otherwise.}
\end{cases}
\\
\Pre_{\perf}[q \step{c_i!m} q'](q,R_p^1, \cdots, R_p^{|\C|}) &=
\begin{cases}
  (p,R_p^1, \ldots, R_p^{i-1}, R_p^i m^{-1}, R_p^{i+1}, \ldots, R_p^{|\C|}) & \textrm{ if } q=r,\\
  \emptyset & \textrm{ otherwise.}
\end{cases}
\\
\Pre_{\perf}\Bigl(\sum_{i\in I} (q_i, R_i^1,\ldots,R_i^{|\C|})\Bigr)
	& = \sum_{i\in I}\sum_{\delta\in\Delta} \Pre_{\perf}[\delta] (q_i,R_i^1,\ldots,R_i^{|\C|}).
\end{align*}
where the notation ``$m R$'' (for concatenation) and ``$R m^{-1}$'' (for
right-residuals) are as in section~\ref{sec-guarded-mucalculus}. For lossy
steps we use
\[
		       \Pre(R) = \Pre_\perf(\Cu(R)).
\]
Clearly, both $\Pre_\perf$ and $\Pre$ are effective operators on regions.

\subsection{Regular model-checking for lossy channel systems}

Surprising decidability results for lossy channel systems is what launched
the study of this model~\cite{finkel94,abdulla96b,cece95}. We reformulate
several of these results as a direct consequence of
Theorem~\ref{th-guarded-effective}, before moving to new problems and new
decidability results in the next sections. Note that our technique is
applied here to a slightly different operational semantics (cf.\
footnote~\ref{STEP-fn}) but it would clearly apply as directly to the
simpler semantics.

\paragraph*{\bf Reachability analysis.}
Thanks to Lemma~\ref{lem-pre-guarded}, the co-reachability set can be
expressed as a guarded $L_\mu$ term:
\begin{equation}
\label{eq-pre-star}
      \Pre^*(V) = \mu X. V \cup \Pre(X) = \mu X. V \cup \Pre(\Cu(X)).
\end{equation}
\begin{corollary}
For regular $V\subseteq\Conf$, $\Pre^*(V)$ is regular and effectively computable.
\end{corollary}

\paragraph*{\bf Safety properties.}
More generally, safety properties can be handled. In CTL, they can be
written $\forall (V_1 \Release V_2)$. Recall that $\Release$, the Release
modality, is dual to Until: a state $\sigma$ satisfies $\forall (V_1 \Release
V_2)$ if and only if along all paths issuing from $\sigma$, $V_2$ always holds
until maybe $V_1$ is visited. Using Lemma~\ref{lem-pre-guarded},
$\rep{\forall (V_1 \Release V_2)}$, the set of states where the safety property
holds, can be defined as a guarded $L_\mu$ term:
\begin{equation}
\label{eq-forall-release}
\rep{\forall (V_1 \Release V_2)}
= \nu X. \bigl(V_2 \cap (\wPre(X) \cup V_1)\bigr)
= \nu X. \bigl(V_2 \cap (\wPre(\Kd(X)) \cup V_1)\bigr).
\end{equation}
\begin{corollary}
\label{coro-AR}
For regular $V_1,V_2\subseteq\Conf$, $\rep{\forall (V_1 \Release V_2)}$ is regular
and effectively computable.
\end{corollary}
Another formulation is based on the duality
between the ``$\forall\Release$'' and the ``$\exists\Until$'' modalities.
\begin{theorem}\cite[sect.~5]{KucSch-TCS}
\label{KS-theorem}
If $f$ is a temporal formula in the
$\Lang(\exists\Until,\exists\Next,\wedge,\neg)$ fragment of CTL (using
regions for atomic propositions), then $\rep{f}$ is regular and effectively
computable.
\end{theorem}
\begin{proof}
By induction on the structure of $f$, using
$\rep{\exists\Next f}\egdef \Pre(\rep{f})$, and the fact that
regions are (effectively) closed under complementation.
\qed
\end{proof}

\paragraph*{\bf Beyond safety.}
Inevitability properties, and recurrent reachability can be stated in $L_\mu$.
With temporal logic notation, this yields
\begin{align*}
\rep{\forall \Diamond V} &= \mu X. \bigl(V \cup (\Pre(\Conf) \cap \wPre(X))\bigr),
\\
\rep{\exists \Box\Diamond V} &= \nu X. \bigl(\mu Y. ((V \cup \Pre(Y)) \cap \Pre(X))\bigr).
\end{align*}
These two terms are not guarded and Lemma~\ref{lem-pre-guarded} is of no
help here. However this is not surprising: firstly, whether
$\sigma\sat\exists\Box\Diamond V$ is undecidable~\cite{abdulla96c}; secondly,
and while $\sigma\sat\rep{\forall \Diamond V}$ is decidable, the set
$\rep{\forall \Diamond V}$ cannot be computed
effectively~\cite{mayr-unreliable}.

\subsection{Generalized lossy channel systems}
\label{ssec-extended-lcs}

Transition rules in LCS's do not carry guards, aka preconditions, beyond
the implicit condition that a reading action $c?m$ is only enabled when
$w(c)$ starts with $m$. This barebone definition is for simplification
purpose, but actual protocols sometimes use guards that probe the contents
of the channel before taking this or that transition. The simplest such
guards are emptiness tests, like ``$p\step{c=\epsilon?}q$'' that only
allows a transition from $p$ to $q$ if $w(c)$ is empty.

We now introduce \emph{LCS's with regular guards} (GLCS's), an extension of
the barebone model where any regular set of channel contents can be used to
guard a transition rule. This generalizes emptiness tests, occurrence tests
(as in~\cite{OuaknineWorrell06}), etc., and allows expressing priority
between rules since whether given rules are enabled is a regular condition.

Formally, we assume rules in $\Delta$ now have the form $p
\step{G:\op} q$ with $p,q,\op$ as before, and where $G$, the guard,
can be any regular region. The operational semantics is a expected:
when $\delta=p\step{G:\op}q$, there is a perfect step
$\sigma\step{\delta}_\perf\theta$ iff $\sigma \in G$ and $\theta$ is
obtained from $\sigma$ by the rule $p \step{G:\op} q$ (without any
guard). Then, general steps $\sigma\step{\delta}\rho$ are obtained
from perfect steps $\sigma\step{\delta}_\perf\sigma'$ by message
losses $\rho\sqsubseteq \sigma'$.

\paragraph*{\bf Verification of GLCS's.}
For GLCS's, $\Pre$ and $\Post$ are effective monotonic
regularity-preserving operators as in the  LCS case since
\begin{align*}
	   \Pre[p\step{G:\op}q](R)&=G\:\cap\:\Pre[p\step{\op}q](R),	\\
	  \Post[p\step{G:\op}q](R)&=\Post[p\step{\op}q](G\cap R).
\end{align*}
Observe that Lemma~\ref{lem-pre-guarded} holds for GLCS's as well, so that
Equations \eqref{eq-pre-star} and \eqref{eq-forall-release} entail a
generalized version of Theorem~\ref{KS-theorem}:
\begin{theorem}
For all GLCS's $\L$ and formulae $f$ in the
$\Lang(\exists\Until,\exists\Next,\wedge,\neg)$ fragment, $\rep{f}$ is
regular and effectively computable.
\end{theorem}



\section{Solving games on lossy channel systems}
\label{sec-turn-based}

In this section, we consider turn-based games on GLCS's where two players,
$A$ and $B$, alternate their moves. Games play a growing role in
verification where they address situations in which different agents have
different, competing goals. We assume a basic understanding of the
associated concepts: arena, play, strategy, etc. (otherwise
see~\cite{Graedel-Thomas-Wilke-LNCS}).

Games on well-structured systems have already been investigated
in~\cite{abdulla2003b,raskin03,raskin05}. The positive results in these
three papers rely on ad-hoc finite convergence lemmas that are special
cases of our Theorem~\ref{th-guarded-effective}.

\subsection{Symmetric LCS-games with controllable message losses}
\label{ssec-symmetric-games}

We start with the simplest kind of games on a GLCS: $A$ and $B$ play in
turn, choosing the next configuration, i.e., picking what rule
$\delta\in\Delta$ is fired, and what messages are lost.

Formally, a \emph{symmetric LCS-game} is a GLCS $\L=
(Q_A,Q_B,\C,\Mess,\Delta)$ where the set of locations $Q=Q_A \cup Q_B$ is
partitioned into two sets, one for each player, and where the rules ensure
strict alternation: for all $p \step{G:\op} q \in \Delta$, $p\in Q_A$ iff
$q\in Q_B$. Below, we shortly write $\Conf_A$ for
$Q_A\times{\Mess^*}^{|\C|}$, the regular region where it is $A$'s turn to
play. $\Conf_B$ is defined similarly. Strict alternation means that the
arena, $\LTS_\L$, is a bipartite graph partitioned in $\Conf_A$ and
$\Conf_B$.

\paragraph*{\bf \em Reachability games.}
Reachability and invariant are among the simplest objectives for
games. In a reachability game, $A$ tries to reach a state in some set
$V$, no matter how $B$ behaves. This goal is denoted $\Diamond V$. It
is known that such games are determined and that memoryless strategies
are sufficient~\cite{Graedel-Thomas-Wilke-LNCS}. The set of winning
configurations for $A$ is denoted with $\ex{A} \Diamond V$, and can be
defined in $L_\mu$:
\begin{equation}
\label{eq-reachgame}
\ex{A} \Diamond V = 
	\mu X.\Bigl[ V \cup \bigl[\Conf_A \cap \Pre(X)\bigr] 
	               \cup \bigl[\Conf_B \cap \wPre(X)\bigr]\Bigr].
\end{equation}
The first occurrence of $X$ can be made upward-guarded by replacing
$\Pre(X)$ with $\Pre(\Cu(X))$ (Lemma~\ref{lem-pre-guarded}). For the second
occurrence, we can unfold the term, relying on the fixpoint equation
$\rep{\mu X.\phi(X)}=\rep{\mu X.\phi(\phi(X))}$. This will replace
$\Conf_B\cap\wPre(X)$ in \eqref{eq-reachgame} with
\begin{equation}
\tag{+}
\label{eq+}
\Conf_B\cap \wPre\Bigl(V \cup \bigl[\Conf_A \cap
  \Pre(X)\bigr] \cup \bigl[\Conf_B \cap \wPre(X)\bigr]\Bigr).
\end{equation}
Now, the strict alternation between $\Conf_A$ and $\Conf_B$ lets us simplify
\eqref{eq+} into
\begin{equation}
\Conf_B \cap \wPre\Bigl(V \cup	\Pre(X) \Bigr).
\end{equation}
Hence \eqref{eq-reachgame} can be rewritten into
\begin{gather}
\label{eq-reachgame-var}
\tag{\ref{eq-reachgame}'}
\ex{A} \Diamond V = \mu X.\Bigl[ V \cup \bigl[\Conf_A \cap
			\Pre(\Cu(X))\bigr] \cup \bigl[\Conf_B \cap
			\wPre(V\cup \Pre(\Cu(X)))\bigr]\Bigr].
\end{gather}

\paragraph*{\bf \em Invariant games.}
In invariant games, $A$'s goal is to never leave  some set
$V\subseteq\Conf$, no matter how $B$ behaves. Invariant games are
dual to reachability games, and the set of winning configurations
$\ex{A}\Box V$ is exactly $\overline{\ex{B}\Diamond \overline{V}}$.

\paragraph*{\bf \em Repeated reachability games.}
Here $A$'s goal is to visit $V$ infinitely many times, no matter how $B$
behaves. The set of winning configurations is given by the following
$L_\mu$ term:
\begin{equation}
\ex{A} \Box \Diamond V = \nu Y. \ex{A}\Diamond \Bigl[ V \cap
(\varphi_A(Y) \cup \varphi_B(Y))\Bigr],
\end{equation}
where
\begin{align*}
	\varphi_A(Y) & \egdef \Conf_A \cap \Pre\bigl( \Cu(\wPre(\Kd(Y)))\bigr),
\\
	\varphi_B(Y) & \egdef \Conf_B \cap \wPre(\Kd(Y)).
\end{align*}
and where we reuse \eqref{eq-reachgame-var} for $\ex{A} \Diamond [\ldots]$.

\paragraph*{\bf \em Persistence games.}
In a persistence game, $A$ aims at remaining inside $V$ from some moment
on, no matter how $B$ behaves. Dually, this can be seen as a repeated
reachability game for $B$. Note that $\ex{A} \Diamond \Box V \neq \ex{A}
\Diamond (\ex{A} \Box V)$.

\begin{theorem}[Decidability of symmetric LCS-games]
For symmetric LCS-games $\L$ and regular regions $V$, the four sets
$\ex{A}\Diamond V$, $\ex{A}\Box V$, $\ex{A}\Diamond\Box V$, and
$\ex{A}\Box\Diamond V$, are (effective) regions. Hence reachability,
invariant, repeated reachability, and persistence symmetric games are
decidable on GLCS's.
\end{theorem}
\begin{proof}[Sketch] The winning sets can be defined by guarded $L_\mu$
  terms.
\end{proof}
\begin{remark}
There is no contradiction between the undecidability of
$\exists\Box\Diamond V$ and the decidability of $\ex{A}\Box\Diamond V$. In
the latter case, $B$ does not cooperate with $A$, making the goal harder to
reach for $A$ (and the property easier to decide for us).
\qed
\end{remark}


\subsection{Asymmetric LCS-games with 1-sided controlled loss of messages}
\label{ssec-1sided-games}

Here we adopt the setting considered in~\cite{abdulla2003b}. It varies from
the symmetric setting of section~\ref{ssec-symmetric-games} in that only
player $B$ can lose messages (and can control what is lost), while player
$A$ can only make perfect steps. Note that this generalizes games where $A$
plays moves in the channel system, and $B$ is an adversarial environment
responsible for message losses. We use the same syntax as for symmetric
LCS-games.

\paragraph*{\bf \em Reachability and invariant games.}
Let us first consider games where one player tries to reach a regular
region $V$ (goal $\Diamond V$), no matter how the other player behaves.

The configurations where $B$ can win a reachability game are given by:
\begin{align*}
\begin{split}
  \ex{B} \Diamond V
  & = \mu X. V \cup \Bigl(\Conf_B \cap \Pre(X)\Bigr) \cup \Bigl(\Conf_A \cap \wPre_\perf(X) \Bigr)\\
  & = \mu X. V \cup \Bigl( \Conf_B \cap \Pre(\Cu (X))\Bigr) \cup
  \Bigl(\Conf_A \cap \wPre_\perf(V \cup \Pre(\Cu(X))) \Bigr)
\end{split}
\end{align*}
where guardedness is obtained via Lemma~\ref{lem-pre-guarded} and
unfolding.

When we consider a reachability game for $A$, the situation is not so clear:
\begin{equation*}
  \ex{A} \Diamond V = \mu X. V \cup \Bigl(\Conf_A \cap
  \Pre_\perf(X)\Bigr) \cup \Bigl(\Conf_B \cap \wPre(X) \Bigr).
\end{equation*}
Neither Lemma~\ref{lem-pre-guarded} nor unfolding techniques can turn
this into a guarded term. This should be expected since the set
$\ex{A} \Diamond V$ cannot be computed
effectively~\cite{abdulla2003b}.

\begin{theorem}[Decidability of asymmetric LCS-games~\cite{abdulla2003b}]
For asymmetric LCS-games $\L$ and regular regions $V$,	the sets
$\ex{B}\Diamond V$  and $\ex{A}\Box V$ are (effective) regions. Hence
reachability games for $B$, and invariant games for $A$ are decidable on
GLCS's.
\end{theorem}
\begin{proof}[Sketch] Invariant games are dual to reachability games, and
  the winning set $\ex{B}\Diamond V$ is defined by a guarded $L_\mu$
  term.
\end{proof}

\section{Channel systems with probabilistic losses}
\label{sec-prob-lcs}

LCS's where messages losses follow probabilistic rules have been
investigated as a less pessimistic model of protocols with unreliable
channels (see~\cite{Sch-voss,ABRS-icomp,BBS-acmtocl06} and the references
therein).

In~\cite{BBS-acmtocl06}, we present decidability results for LCS's seen as
combining \emph{nondeterministic} choice of transition rules with
\emph{probabilistic} message losses. The semantics is in term of Markovian
decision processes, or $1\frac{1}{2}$-player games, whose solutions can be
defined in $L_\mu$. Indeed, we found the inspiration for $L_\mu$ and our
Theorem~\ref{th-guarded-effective} while extending our results in the MDP
approach to richer sets of regions.

In this section, rather than rephrasing our results on
$1\frac{1}{2}$-player games on LCS's, we show how to deal with
$2\frac{1}{2}$-player games~\cite{chatterje05} on LCS's, i.e., games
opposing players $A$ and $B$ (as in section~\ref{sec-turn-based}) but where
message losses are probabilistic.
\\

Formally, a \emph{symmetric probabilistic LCS-game}
$\L=(Q_A,Q_B,\C,\Mess,\Delta)$ is exactly like a symmetric LCS-game but
with an altered semantics: in state $\sigma\in\Conf_A$, player $A$ selects
a fireable rule $\delta\in\Delta$ ($B$ picks the rule if
$\sigma\in\Conf_B$) and the system moves to a successor state $\rho$ where
$\sigma\step{\delta}_\perf\sigma'\sqsupseteq\rho$ and $\rho$ is chosen
probabilistically in $\Cd(\{\sigma'\})$. The definition of the probability
distribution $\bfP(\sigma,\delta,\rho)$ can be found
in~\cite{Sch-voss,BBS-acmtocl06} where it is called \emph{the local-fault
model}. It satisfies $\bfP(\sigma,\delta,\rho)>0$ iff
$\rho\sqsubseteq\sigma'$ (assuming $\sigma\step{\delta}_\perf\sigma'$).
Additionally it guarantees a \emph{finite-attractor property}: the set of
states where all channels are empty will be visited infinitely many times
almost surely~\cite{ABRS-icomp,BBS-ipl}.

\paragraph*{\bf \em Reachability games.}
Assume $A$ tries to reach region $V$ (goal $\Diamond V$) \emph{with
probability 1} no matter how $B$ behaves. The set $\ex{A} [\Diamond
V]_{=1}$ of states in which $A$ has an almost-sure winning strategy is
given by
\begin{equation}
\label{eq-reach-P1}
\ex{A} [\Diamond V]_{=1} =
 \nu Y. \mu X.	 \left(
	\begin{array}{rl}
	 V \cup & \Bigl[\Conf_A \cap \Pre_\perf (\Cu(X) \cap \Kd(Y))\Bigr]
    \\[.5em]
	   \cup & \Bigl[\Conf_B \cap \wPre_\perf (\Cu(X) \cap \Kd(Y))\Bigr]
	\end{array}
	\right).
\end{equation}
\begin{remark}
Justifying \eqref{eq-reach-P1} is outside the scope of this paper, but we
can try to give an intuition of why it works: the inner fixpoint ``$\mu X.
V\cup \cdots$'' define the largest set from which $A$ has a strategy to
reach $V$ no matter what $B$ does \emph{if the message losses are
favorable}. However, whatever messages are lost, $A$'s strategy also
guarantees that the system will remain in $Y$, from which it will be
possible to retry the strategy for $\Diamond V$ as many times as necessary.
This will eventually succeed almost surely thanks to the finite-attractor
property.
\qed
\end{remark}

\paragraph*{\bf \em Invariant games.}
Assume now $A$ tries to stay in $V$ almost surely (goal $[\Box V]_{=1}$), no
matter how $B$ behaves. Then $A$ must ensure $\Box V$ surely and
we are considering a 2-player game where message losses are adversarial
and could as well be controlled by $B$. This leads to
\begin{equation}
\label{eq-Box=1}
\begin{split}
\ex{A}[\Box V]_{=1}
& = \nu X. V \cap
\Bigl(\bigl[\Conf_A \cap
\Pre_\perf(\Kd(X))\bigr] \cup \bigl[\Conf_B \cap \wPre(X)\bigr]\Bigr)
\\
& = \nu X. V \cap
\Bigl(\bigl[\Conf_A \cap
\Pre_\perf(\Kd(X))\bigr] \cup \bigl[\Conf_B \cap \wPre(\Kd(X))\bigr]\Bigr).
\end{split}
\end{equation}
In \eqref{eq-Box=1}, the subterm $\Pre_\perf(\Kd(X))$
accounts for states in which $A$ can choose  a perfect move that will end
in $\Kd(X)$, i.e., that can be followed by any adversarial message losses
and still remain in $X$. The subterm  $\wPre(X)$ accounts for states
in which $B$ cannot avoid going to $X$, even with message losses under his
control.  $\wPre(X)$ can be rewritten into
$\wPre(\Kd(X))$ thanks to Lemma~\ref{lem-pre-guarded}, so that we end up
with a guarded term.

\paragraph*{\bf \em Goals to be satisfied with positive probability.}
In $2\frac{1}{2}$-player games, it may happen that a given goal can only be
attained with some non-zero probability~\cite{chatterje05}. Observe that,
since the games we consider are determined~\cite{martin98}, the goals
$[\Diamond V]_{>0}$ or $[\Box V]_{>0}$ are the opposite of goals asking for
probability 1:
\begin{xalignat*}{2}
\ex{A}[\Diamond V]_{>0} &= \overline{\ex{B}[\Box \overline{V}]_{=1}},
& \quad
\ex{A}[\Box V]_{>0} &= \overline{\ex{B}[\Diamond \overline{V}]_{=1}}.
\end{xalignat*}

\begin{theorem}[Decidability of qualitative symmetric probabilistic LCS-games]
For symmetric probabilistic LCS-games $\L$ and regular regions $V$, the
sets $\ex{A}[\Diamond V]_{=1}$, $\ex{A}[\Diamond V]_{>0}$, $\ex{A}[\Box
V]_{=1}$, and $\ex{A}[\Box V]_{>0}$ are (effective) regions. Hence
qualitative reachability and invariant games are decidable on GLCS's.
\end{theorem}
\begin{proof}[Sketch]
These sets can be defined by guarded $L_\mu$ terms.
\qed
\end{proof}

\section{Conclusion}
\label{sec-concl}

We defined a notion of upward/downward-guarded fixpoint expressions that
define subsets of a well-quasi-ordered set. For these guarded fixpoint
expressions, a finite convergence theorem is proved, that shows how the
fixpoints can be evaluated with a finite number of operations. This has a
number of applications, in particular in the symbolic verification of
well-structured systems, our original motivation. We illustrate this in the
second part of the paper, with lossy channel systems as a target. For these
systems, we derive in an easy and uniform way, a number of decidability
theorems that extend or generalize the main existing results in the
verification of temporal properties or game-theoretical properties.

These techniques can be applied to other well-structured systems, with a
region algebra built on, e.g., upward-closed sets. Admittedly, many
examples of well-structured systems do not enjoy closure properties as nice
as our Lemma~\ref{lem-pre-guarded} for LCS's, which will make it more
difficult to express interesting properties in the guarded fragment of
$L_\mu$. But this can still be done, as witnessed
by~\cite{raskin03,raskin05} where the authors introduced a concept of
$B$-games and $BB$-games that captures some essential closure assumptions
allowing the kind of rewritings and unfoldings we have justified with
Lemma~\ref{lem-pre-guarded}.







\end{document}